\documentclass{llncs}

\usepackage{url} % usato per qualche links 
\usepackage{epsfig}

\usepackage{graphicx}
\usepackage{comment}
\usepackage{subfigure}
\usepackage{wrapfig}
\usepackage{color} % for develop purposes
\makeatletter
\newif\if@restonecol
\makeatother

\usepackage[linesnumbered]{algorithm2e}

\newcommand{\fs}{\emph{FS}}

\newcommand{\genoma}{\emph{genome}}
\newcommand{\fsgenoma}{\emph{File System Genome}}
\newcommand{\fsg}{\emph{FSG}}

\usepackage{hyphenat}   % per le parole a capo
\usepackage{amsmath}
\usepackage{amssymb}
\usepackage{mathtools}

\begin{document}

\title{Mapping the File Systems Genome: rationales, technique, results and applications}

\author{Roberto Di Pietro\inst{1} \and Luigi V. Mancini\inst{2} Antonio Villani\inst{1} \and Domenico Vitali\inst{2}  }

\institute{Universit\`{a} di Roma Tre, Dipartimento di Matematica
\email{\{dipietro,villani\}@mat.uniroma3.it}
\and
Universit\`{a} di Roma La Sapienza, Dipartimento di Informatica
\email{\{mancini,vitali\}@di.uniroma1.it}
}

\maketitle

\begin{abstract}
This paper provides evidence of a feature of Hard-Disk Drives (HDDs), that we
call \fsgenoma. Such a feature is originated by the areas where (on the HDD) the
file blocks are placed by the operating system during the installation
procedure. 
It appears from our study that the \fsgenoma\ is a distinctive and unique
feature of each individual HDD. In particular, our extensive set of experiments
shows that the installation of the same operating system on two identical
hardware configurations generates two different File System Genomes. Further,
the application of sound information theory tools, such as min entropy, show
that the differences between two 
\fsgenoma\ are considerably relevant. 
The results  provided in this paper constitute the scientific basis for a number
of applications in various fields of information technology, such as forensic
identification and security. 
Finally, this work also paves the way for the application of the highlighted
technique to other classes of mass-storage devices (e.g. SSDs, Flash memories).
\end{abstract}

\section{Introduction\label{sec:intro}}
This paper reports evidence of an apparently surprising phenomenon that occurs
in storage devices. In particular, this paper takes into consideration
computer systems that are perfectly identical both from the hardware point of
view (storage devices, motherboard, CPU, RAM, etc..) and software point of view
(operating system and applications), and shows that 
the locations occupied by the file's blocks 
differ greatly between any two
storage devices of such identical computer systems. Such a distribution of file
blocks has an entropy even greater if one analyses different computer system in
terms of hardware and software. In other words, let us assume that we install on a
computer C a given operating system, and let $P$ be the set of locations
occupied by the file blocks at the end of this operating system installation.
After formatting the storage device and having installed  the same operating system 
on the same computer C, we observed that the new set of locations $P'$ occupied by
the file blocks at the end of the installation process
significantly differs from $P$.
The complete set of the locations of the file blocks on a given storage device
are called in this paper \fsgenoma\ (\fsg). It appears from our study
that the \fsgenoma\ is a distinctive and unique feature of each
individual storage device. 

The following sections discuss the underlying physical and logical phenomenons
originating 
this feature, and present an extensive experimental campaign to support our
observations. In particular, these first time measured pieces of evidence have
been analyzed  in a rigorous way by using information theory techniques. 
In particular, our results demonstrate that the installation of the same
operating system software on two identical hardware configurations generate two
significantly  different \fsg. \\
The concept of \fsg\ seems to 
be applicable to many fields of computer science,which include for example 
Forensic Identification. The Forensic Identification of electronic devices is
the ability to identify model, configuration, and other characteristics of an
electronic device in a court case. The characteristics that allow the unique
identification of a device are called device signatures. Such an ability provides a
valuable aid for law enforcement and intelligence agencies, especially in the
presence of illicit activities. In general, there are many scenarios in which
the Forensic Identification can be decisive. For example, it may be used to
confirm whether digital photographs have been acquired using a specific
camera~\cite{securedigitalcamera}, where the noise characteristics in a digital
image is used as camera's signature; or it may be useful to confirm that a
document has been printed out using a specific printer~\cite{survey-forensics}.
Similarly, the concept of \fsg\ can be used to develop a scheme to verify the
ownership of a particular storage device.
For example, consider the following scenario: suppose that the IT security
officer of an organization has to verify that a lost storage device, just found
again, belongs to the organization. The aim of the adversary is to forge the
content of a storage device in such a way that it is considered as belonging to
the organization, so as to have it introduced in the IT infrastructure of the 
opponent---for instance, this device carrying a stuxnet-like worm. Assume that
the security officer has computed previously the
\fsg\ on the read-only files of his storage device, keeping it in a safe place. If
the \fsg\ of the found storage device corresponds with the one the IT security
officer had guarded, then he can demonstrate the property of the item. Note
that since the IT security officer computes the \fsg\ only on read-only files, the
\fsg\ remains valid and unchanged, even if the IT security officer modifies the
storage device after calculating the \fsg. A simple solution based on computing
the hash of the entire storage device is unsatisfactory because two storage devices
with the same data have the same hash, 
while they have different  \fsg, as it will be shown in the following.
In addition, \fsg\ can be a useful building block for
authentication protocols as well. Consider for instance the following scenario:
the security policies of a Government Organization prevents the employees from bringing electronic devices belonging to the organization (namely internal
devices) outside the organization building. At the same time,  such security
policies allow an employee to bring his own electronic devices (namely external
devices), such as laptop or smart phones, inside the organization's building.
The Government Organization can leverage \fsg\ to reliably distinguish between
internal and external devices.

\paragraph{Contributions} 

Our results evidence a distinctive and unique feature of each
individual storage devices, called \fsgenoma\ (\fsg). In particular, the
main contributions of this paper are:
\begin{enumerate}
  \item We put forward the definition of \fsg, which can be used to
distinguish two devices even if they use the same configurations (hardware
and software) and store the same data. To the best of our knowledge, this
feature has not been considered before.
  \item We devise an algorithm to map the \fsg\ compliant with ext4, one of the
most spread file systems used in both server and client workstations and we
collect a huge set of real installations of the Debian GNU/Linux (more than 
$12,000$ installations of such operating system).
  \item We evaluate the \fsgenoma\ by investigating the entropy
and min-entropy values, that characterize the diversity of the installations
as stochastic process %and we measure the distance between installations
\end{enumerate}

Our results prove that an host can be reliably identified based on the
\fsgenoma,  showing a straightforward application to the forensics
identification domain.
Further, we believe that the \fsgenoma\ can be used as a building block for
authentication protocols and, in general, to enforce other security properties.
\begin{comment}
Based on these observations, we define the \fsgenoma\ as the list of block locations used by a set 
of files. 
\begin{definition}[File System Genoma (FSG)]\label{def:fs-topology}
Let $\mathbb{F}$ be a File System hierarchy; 
The File System Genoma, $\mathbb{G}_{\mathbb{F}}$, is a matrix 
where each row $i$ represents the list of locations used to store the data blocks of the
file $f_i$ belonging to $\mathbb{F}$.
\end{definition}
The surprising results exposed in this paper highlight that \fsgenoma\ presents
a large amount of ramdomness and can be used as \textit{signature} of operating
systems installations. 

\end{comment}

\paragraph{Roadmap}
The paper is organized as follows: Section~\ref{sec:topology} 
describes the causes of the non-determinism of the file block locations;
it investigates the relevant aspects in the architecture of a modern 
operating systems focusing on the Linux kernel.
Section~\ref{sec:evaluations} describes the environment and testbed in which
our experiments have been performed and the evaluations of the proposed
fingerprinting technique. Section~\ref{sec:related} surveys related work,
while Section~\ref{sec:conclusions} concludes the paper 
outlining future directions. 

\section{The non-determinism of the file block locations\label{sec:topology}}
This section investigates the software and hardware components that are involved
in the choice of the block addresses selected during a write operation. Due to
the unpredictable interleaving of these components, the overall result is the
introduction of a certain level of randomness in file allocation, that
originates the \fsg\ diversity we measured in our experiments. This section also
provides a description of the main features of the file system we based our
experiments upon, and survey how the related structures influence the \fsg.

In this paper we focus on Hard Disk Drives (HDDs): a type of storage device that
stores data on rotating magnetic disks. The most important components of HDDs
are: the \textit{mechanical arm}, the \textit{head} and the disk. These
components are involved during every write operation; to perform a write, the
arm places the head in the right position; then, the head writes the data on the
rotating disk. The arms placements is the most time consuming operation, referred as 
\textit{seek time}.

To reduce the seek time it is convenient to limit the movements of the arms. 
One of the kernel components that is involved in the seek-time optimization is
the File Systems (\fs). Indeed, whenever a \emph{user-space} process invokes 
the \emph{write()} system call, the data are copied in kernel space, splitted into 
blocks and prepared for the \emph{physical} write~\cite{understanding}. 
When a \emph{write()} system call returns, data are just placed in a \emph{write
cache} that  resides in RAM. As such, there are no block locations that are
marked as occupied when the \emph{write()} system call returns to the caller.
Through this mechanism,
multiple physical write requests can be aggregated, reducing the mean \textit{seek time}.
The physical write on the storage device is performed by
the kernel in predetermined time intervals through a specific operation called
\emph{sync}. The choice of the locations where the blocks has to be placed is
performed by the \emph{sync} operation, on the basis of the free blocks and the
blocks present in the \emph{write cache}.

Based on these observations, we define the \emph{File System Genome} as the list of block
locations used by a set of files. 
\begin{definition}[File System Genome (FSG)]\label{def:fs-topology}
Let $\mathbb{F}$ be a File System hierarchy; 
The File System Genome, $\mathbb{G}_{\mathbb{F}}$, is a matrix 
where each row $i$ represents the list of locations used to store the data blocks of the
file $f_i$ belonging to $\mathbb{F}$.
\end{definition}
The surprising results exposed in this paper highlight that the \fsgenoma\ 
presents a large amount of randomness may represent a distinctive and unique 
feature of each individual storage device.

In order to explain this phenomenon, consider the installer of a generic operating system. 
In Figure 1, during the installation process several files are written to the
disk. 
\begin{figure}
\centering
  \includegraphics[width=.7\textwidth]{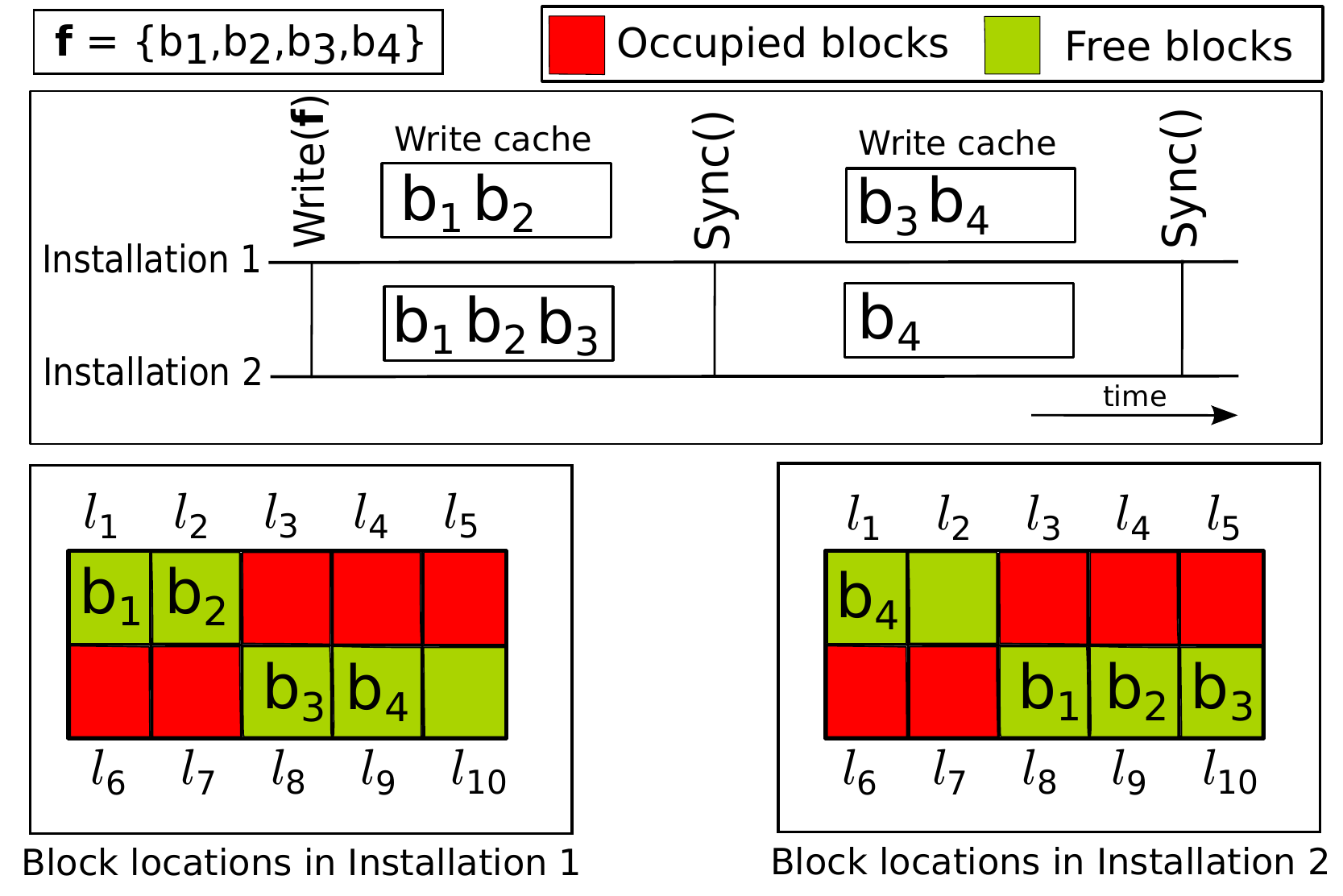}
  \label{fig:block_assignment}
  \caption{The writings of file $f$ in two independent installations. The
blocks in the write queue influence the choices of the physical locations.}
\end{figure}
We take as a reference a generic file $f$, which is composed by four blocks 
($b_1,b_2,b_3,b_4$). When the sync is invoked by the kernel, a different set of blocks
is present within the \emph{write cache} of the two installations. During
installation $1$ only $b_1$ and $b_2$ are in the write cache at the time of
sync. In this case, the best locations (i.e. the locations that minimize the
seek time) are all the positions with at least two consecutive free blocks, that
is the locations $l_1,l_2$. Whereas, during  installation $2$, $b_1,b_2$ and
$b_3$ are in the \emph{write cache} at the time of the sync. In this case there
is only one possible solution: $l_8,l_9,l_{10}$. From this point onwards, the
locations selected for the subsequent physical writes during installation $1$
diverge from the one selected by the installation $2$. Please note, 
when the installer invokes the write system call for $f$, the lists of free and
occupied blocks are identical for the two installations, nevertheless that the
described operating system optimization mechanisms can cause significative
variations in the choices of the block locations. Further, Figure 1 describes
only one example whereas there are many possible cases that depend on several
factors such as the interleaving of the processes.

In two consecutive operating system installations, a different set of
blocks can be present in the write cache for the same file. In
addition, if we consider that the execution of the write and the sync
operations are totally asynchronous and can have unpredictable
interleavings, one can conclude that the overall effect is a
non-determinism in the choice of the block location during each
installation. 
Also, we have assumed for simplicity in Figure 1 that the sync operations occur at the
same time during two different installations, however in reality such sync operations are unrelated, 
hence the degree of non-determinism is even higher as will be demonstrated in the following.

The following subsection provides a more detailed description of the
unpredictable interleaving of \fs\ operations in the case of the Linux kernel. 

\subsection{Linux I/O architecture for storage devices\label{sec:unixfs}}
Linux supports multiple file system types through a multi-tiered
software architecture. 
In this section, we describe the kernel components involved during an I/O
operations, focusing on the most widespread file system in the Linux environment: the
Extended File System version $4$ (\emph{ext4}). 

\subsubsection{Virtual File System\label{sec:vfs}}
The Virtual File System (VFS) is an intermediate layer between the file system modules
(e.g. \textit{ext4}) and the kernel system calls. 
The first advantage of using the VFS is that different file systems can be
managed in a similar way by the kernel.
The interfaces exposed by the VFS are wrappers, that
forward a function to the \fs\ specific implementation. 
The abstraction provided by the VFS allows a user-level developer to invoke the I/O 
system calls (e.g. \emph{read() and write()}) without worrying about the
specific underlying \fs\ . The VFS implements a common file model and all the \fs\ 
drivers must comply with it.\\
The VFS splits large I/O requests into chunks of fixed-size and forces single-block 
allocations without providing any information about the pending 
operations on that file to the \fs\ implementation. Due to a different
interleaving of the processes, the number of blocks that has been forwarded 
to the lower layers at the sync time, can vary across different execution of 
the installation procedure. 
As such, the VFS influences which blocks are marked as free and which blocks
are present in the write cache when the kernel invokes the sync operation. 

\subsubsection{The Ext4 file system\label{sec:ext4}}
\emph{Ext4} inherits several features from  its previous versions; for this
reason in the following we provide an overview of \emph{ext2} and \emph{ext3}.

Ext2 and ext3 introduced several new features to improve the performance and the 
resilience to disk failures of the \fs.  
In order to limit the internal fragmentation, they organize the blocks 
into groups. Each group covers adjacent tracks and includes both data
and meta-information. In details, each block group contains: i) a copy of the
file system's \textit{superblock}; ii) a \textit{group descriptors}; iii) the
block group \textit{bitmap}, namely a structure that takes into account the
free blocks in the group; iv) the \textit{inode data structures} (bitmap and
table); and, v) the data blocks. 
The ext2 and ext3 file systems preallocate exceeding data blocks to regular files before they
are actually used. Such a strategy reduces fragmentation when  the file size increases,
since the data are allocated in adjacent location.

Ext4 is the last version of the family~\cite{ext4}. Ext4
supports larger file and larger partitions with respect to the previous
versions, and introduces several new features. However, for the sake of this
work, we focus on two of these features: i) the \textit{delayed allocation}; 
and, ii) the \textit{multiple allocation}.
\emph{Delayed allocation}, defers block allocations from the time of the 
write system calls to the time of \emph{sync}. 
This feature brings several advantages: it increases the opportunity to combine 
many block allocation requests into a single request reducing fragmentation and 
saving CPU cycles; further, it avoids unnecessary block allocation 
for short-lived files. At the time of the write system call, ext4 estimates 
only the required number of blocks to reserve (marking it with the \emph{BH\_DELAY} flag) 
checking that enough free blocks are available in the file system to satisfy the write. 
Later, when the sync operation starts, all the dirty blocks (i.e. marked as BH\_DELAY)
are processed and clustered according to the \emph{logical locality} that is 
established by the \textit{multiple allocation} component. This component attempts 
to accommodate the requirements of both large and small files; in details, for small files 
ext4 uses a per-CPU locality policy which is shared by all allocations under the same CPU; 
this strategy allows to keep small files close to each other. In contrast, the allocation request 
of large files is performed using a per-file locality policy; as such, ext4 
maintains an in-memory preallocation range for each file, and uses that to solve 
the fragmentation issues caused by concurrent allocations.
With respect to the ext3 block allocation policy (single-block allocation), ext4
provides better performance for both small and large files~\cite{ext4}.
These features of ext4 may contribute to increment the \fs\ diversity; 
on the one hand, the per-CPU locality binds a write operation to the 
scheduling order, that cannot be considered deterministic. On the other
hand, for two independent installations, a different set of blocks may be present in the
write cache for the same file. Such diverse settings could lead to different allocation
decisions; in particular, if only few blocks of a large file are present in the write cache, 
at the time of the sync, then the  \textit{multiple allocation} could wrongly 
exchange a large file for a small one. Due to the differences in the locality policies 
between large and small files, in ext4 this scenario cause differences in the
the block locations, influencing the \fsg.

Ext4 introduced several improvements in the block group management (e.g. flex\_bg \cite{ext4}), 
however, as in the previous versions, also in ext4 the top level directories are
always spread in different groups. In the other cases, ext4 behaves as follows: 
if the new inode is a directory, then a forward search is made for a block group with both
free space and a low directory-to-inode ratio; if that fails, then of
a \emph{random group} is returned. 
For other inodes, a forward search is performed from the parent 
directory's block group to find a free inode 
(for more details please refer to the file \textit{fs/ext4/ialloc.c} in the
version 3.2.39 of the linux sources).  
Under these circumstances, during the installation procedure, the choices 
of the groups of the first files is demanded to a random function. Further,
due to the \emph{delayed allocation} and \textit{multiple allocation}, also
if the same block group is chosen in two different installations, there 
are differences in the obtained \fsg.

\subsection{Hardware faults\label{sec:hardware}}

This subsection describes how the status of an HDD may influence the \fsg.

Some physical blocks of an HDD can be marked as damaged by the widely used 
Self-Monitoring, Analysis and Reporting Technology system (SMART)
\footnote{http://en.wikipedia.org/wiki/S.M.A.R.T.}.

The SMART controller provides a way to measure drives' characteristics at
the firmware layer, by collecting the functional statistics of the HDD, e.g. track-seek
retries, read errors, write faults, reallocated sectors, head fly height, and
environmental temperature. Once the SMART recognizes an high error rate for some 
physical blocks, that blocks are marked as unusable. 

Many factors can causes damages to an HDD. Damaged blocks, generated by manufacturing errors,  
are already present in the devices when they leaves the factory. 
In such case, the HDD owner will never uses
the bad blocks, in some circumstances it may modify
the disk's geometry.
More in general, an HDD can receive shocks (physical or electrical) and some blocks may be damaged.
An HDD can also have one or more \emph{bad blocks}. These blocks are handled
in a different way by the SMART. The disk, at first, try to recover the information;
next, it moves the recovered information to one or more spare blocks. The 
original block will be marked as 'bad' and will no longer be addressable. The spare
blocks are a limited resource.

Hardware faults, device lifetime, custom configurations or usage patterns
contribute to the definition of the \fsg.

\section{Experiments and discussions\label{sec:evaluations}}

In order to provide a reliable support to our solution, we performed an
extensive experimental campaign, described in detail and discussed in the
following. 

\begin{comment}
Figuring to implement a security protocol on the top of the genome, we should 
ensure how it is unpredictability by an adversary, in other terms, how hard is
to compute a collision by implementing the genome algorithm with a lot of
different inputs. 
This aspect is relevant for many context: in the simplest case, it may be used
to define the minimum amount challenges/response pairs exchanged in a generic
protocol between tester and verifier; in more complex instances, these analysis
are fundamental to obtain property of Physical Unclonable Function,
PUF~\cite{formalpropertypuf}.
Other works use the same approach for our analysis. In~\cite{accellerometro},
authors propose to use the accellerometer of a device for security purposes, in
details they use it for the purposes of true random number generation. 
\end{comment}

\subsection{Test-bed and settings\label{sec:exp-env}}

In this section we describe the environment for our experiments and the strategy
followed to obtain our results. %We performed the evaluation on a statistically 

Our testbed includes the GNU/Linux OS Debian (version 6.0.6,
code-name \textit{squeeze}) and the \emph{Ext4} file system implemented on the Linux
Kernel (version $3.2.32-1$). We choose this configuration due to its open-source
nature and its wide spread diffusion; indeed, the availability of the source
code allowed a deeper investigation on the causes of the randomness in the
\fsg. \\ %of the files allocated by operating system as described in Section~\ref{sec:topology}.\\
Usually, the installation procedure requires a large amount of time, since it
is an interactive process and often it involves slow-access devices (e.g. CD-Rom
or DVD). In order to mitigate these aspects, 
we apply two strategy. On the one hand, we exploit the \textit{preseed} feature
of the
Debian installer; the preseeding strategy simply provides a way to set answers
to questions asked during the installation process. In such way, there is no
need for a user monitoring the installation procedure since this task becomes fully
automated. 
On the other hand, in order to reduce the time required to complete the
installation procedure, we copied the entire Free Software distribution of
Debian GNU/Linux on an USB drive, and used it as a source media for the actual
operating system installation.

The \textit{preseed} feature requires the definition of a configuration file
\textit{preseed.cfg} located in the root of the \textit{initrd} (initial
ramdisk). The answers written in this file are read during the loading stage
and they define the behaviour of the Debian software. In such a way, we can
define how the installer works in advance; in details:
\begin{itemize}
  \item \textit{Environment.} The keymap, the language the country and the time zone
  are set to US; the hardware provides an Ethernet card
configured by the dhcp server; the installer creates two users (root and guest)
  \item \textit{Disk settings.} We do not use LVM or RAID technologies, but
we required a regular installation: we use the entire disk and create the
swap and the root partitions using the \textit{atomic} mode in the 
preseed configuration where all files are placed in single partition. 
Since it is the default setting for the installer, we
claim this parameters are the common ones used by the GNU/Linux users.
  \item \textit{Post installation procedure.} In addition, we have used the Post
installation procedure to gather all the data and the statistics needed for our
study. In particular, when each installation is completed, a script (named
osfingerprint.sh) is started. This script fetches and saves the data that will
be used in the subsequent data analysis. The post installation procedure
collects the list of the blocks locations used by each file of the file
systems.Each reboot triggers a new installation.
\end{itemize}
The post installation procedure gets the list of
the blocks locations used by each file of the file systems. These data have
gathered through the tool \textit{debugfs}, a linux tool which can be used to examine 
and change the state of an ext2, ext3, or ext4 file system. All the data collected after
each installation are sent to a remote repository.

All the experiments have been performed in the Computer Lab of the Computer
Science Department of the Sapienza University of Rome. 
This laboratory is equipped with 28 workstations, having all the same hardware
configuration: Intel(R) Pentium(R) 4 CPU $@3.00GHz$, $2GB$ of volatile memory
(DDR $533Mhz$) and $81GB$ of Hard Disk. We stress the fact that all the workstations have the same
hardware configuration, being physically indistinguishable from each other. Moreover,
the installed CPU does not provide Hyper-Threading support. It is important to
recall that a multi-threaded environment may increase the non-determinism
increasing the diversity between the installations. In this context,
our testbed represents a worst case for our solution.

The experiments required about two weeks, and were carried out during the
Christmas holidays. We collected and analyzed about $12,000$ installations. 
We claim this is a relevant data set providing a reliable support for the assessment 
of the \fsg\ . To the best of our knowledge, this is the first dataset harvested 
from real operating system installations.

\subsection{Installation corpus\label{sec:corpus}}
In our experiments, we created two distinct dataset: i) the \textit{standard} dataset, that represents
the installations with the basic packages; and, ii) the \textit{full}
dataset, that represent the installations enriched with additional software. In particular, the
\textit{standard} installation is composed by $26,389$ files whereas the
\textit{full} installation contains $121,587$ files.
Table~\ref{tab:comp-install} reports a high-level comparison between the
\textit{full} and the \textit{standard} installations.

\begin{table}[!h]
  \begin{center}
    \begin{tabular}{| c | c | c |}
    \hline
    \textbf{Type}  & \textbf{Standard} & \textbf{Full}  \\
    \hline
    \textbf{Total Num. Files} & 26389 & 121587 \\
    \hline
    \textbf{Total Num. Blocks} & 164360   & 873861 \\
    \hline
    \textbf{Required Disk space} & 642 MB & 3413 MB \\
    \hline
    \end{tabular}
      \caption{A high-level comparison between the \textit{full} and the
\textit{standard} installations.\label{tab:comp-install}.} 
    \end{center}
\end{table}  

In details, the \textit{standard} installation includes a desktop environment
(Gnome 3) and common tools (e.g. a rich text editor like Libreoffice, graphics
software, shell script etc.).
Further, the \textit{standard} installation includes some server environment
software, e.g. the web-server (the Apache2 developed by the Apache Software
Foundation), print-server (Cups systems), the DNS-Server (The ISC Bind
application), file-server (Samba), mail-server (Exim Internet Mailer, namely the
default choice of the Debian installer) and the SQL-database (Oracle MySQL
server). 

\begin{table}[!h] 
\begin{center}
    \begin{tabular}{| p{3cm} |  p{3cm}  |  p{3cm}  |}
    \hline
    \textbf{Number of blocks}  &  \textbf{Standard}  & \textbf{Full}  \\
    \hline
    \hline
    [0 - 9] & $91,35\%$ (24108) & $90,92\%$ (110550)\\
    \hline
    [10 - 99] & $7,94\%$ (2096) & $8,31\%$ (10110)\\
    \hline
    [100 - 499] & $0,58\%$ (154) & $0,63\%$ (768)\\
    \hline
    [500 - 999] & $0,09\%$ (24) & $0,07\%$ (93)\\
    \hline
    [1000 - 1999] & $0,01\%$ (3) & $0,028\%$ (35)\\
    \hline
    [2000 - 3999] & $0,015\%$ (4) & $0,017\%$ (21)\\
    \hline
    \hline
    more than 4000 & $0\%$ (0) & $0,012\%$ (7) \\
    \hline
    \end{tabular}
  \end{center}
  \caption{Percentage of files composed by a certain
number of blocks. \label{tab:comp-blocks} Between the brackets is reported the
corresponding absolute value.}
\end{table}

In part of the experiments, we consider only the first block allocated for a file.
We claim that this aspect does not adversely affect our results since the majority 
of the files of an operating system is stored in just few blocks as 
showed in table~\ref{tab:comp-blocks}.

\subsection{Evaluations\label{sec:experiments}}
In order to introduce our full experiments, we have first studied the contribution of three specific files 
to the characterization of the \fsg .

\subsubsection{Single file analysis and representation of the \fsgenoma\ \label{sec:exp-singlefiles}}
We performed a punctual analysis with the aim to explore how single files can contribute 
to the diversity of the File System Genome.

\begin{figure}[!h]
  \centering  
  \includegraphics[angle=-90,width=.65\textwidth]{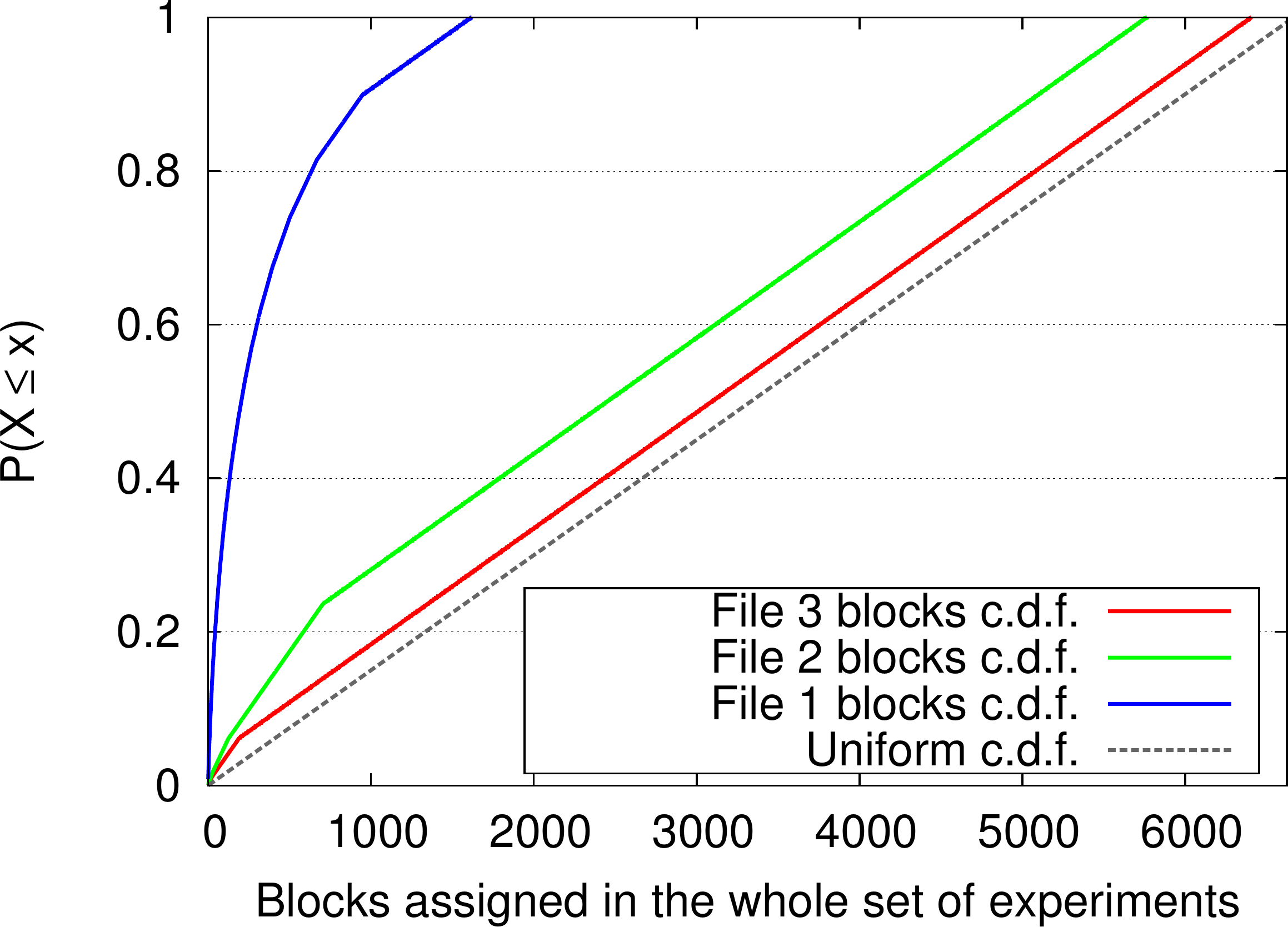}
  
  \caption{The Cumulative Distribution Function (c.d.f.) for three files\label{fig:blocks_cdf}}
\end{figure}

We have chosen three specific files that are part of the standard installation of Debian GNU/ Linux OS. 
The files considered occupies a single data block and are written in the following order:
\textit{File 1} before \textit{File 2} and \textit{File 2} before \textit{File 3}. 
Figure~\ref{fig:blocks_cdf} shows the Cumulative Distribution Function (c.d.f.) related to 
the three random processes after $6,625$ installations of these three specific files.
In the picture, the dotted line represents the uniform c.d.f, namely the ideal distribution where 
the block allocator choose a different block with the same probability in each installation. 
The other lines report the measured c.d.f. of the considered files. This figure shows that the random
process of \textit{File 1} appears more predictable than the random process of \textit{File 3}. Indeed, 
the random process of File 1 does not appear to be uniform: a file block is stored with probability one 
in the first $1,500$ locations. However, the c.d.f of File 2 and File 3 tend to approximate the 
ideal uniform distribution represented by the dotted line. This is due to the fact that the initial 
state of the installer procedure, is equal for all the installations. As the installation proceeds, 
the entropy of the random process gradually increases, getting close to the uniform distribution.
\begin{comment}
In order to provide  an overview of such behaviour, we wrote $121,587$  files for
$6,625$ rounds, tracking the blocks where the files were placed in each round. 
We modeled the choice made by the block allocator as a probability distribution;
Figure~\ref{fig:blocks_cdf} shows the Cumulative Distribution Function (cdf) of
this process for three \emph{small files} (i.e. smaller than 4096 bytes).
The dotted line represents the ideal case where the block allocator choose a
different block in each round (i.e. uniform distribution).
As the cdf shows, each plotted file is characterized by a specific distribution;
indeed, the number of different blocks assigned to File 1 are comparable to the
uniform distribution whereas the blocks assigned to File 3 are more
predictable.
\end{comment}
\begin{comment}
to enable the design and the implementation of a strong and easy-to-use 
random data accumulator and generator which requires no specialized hardware.
dire che e' una fonte di entropia difficile da influenzare

Since the target block for a write operation depends on the free 
blocks available at the time of the operation request, each component actively
influences 
also the location (i.e. the target block) \emph{where} the data are placed. 

As such, the exact location where a data should be placed in the storage
device is unpredictable in advance.

\end{comment}

\begin{figure}[!h]
  \centering 
  \includegraphics[angle=-90,width=.65\textwidth]{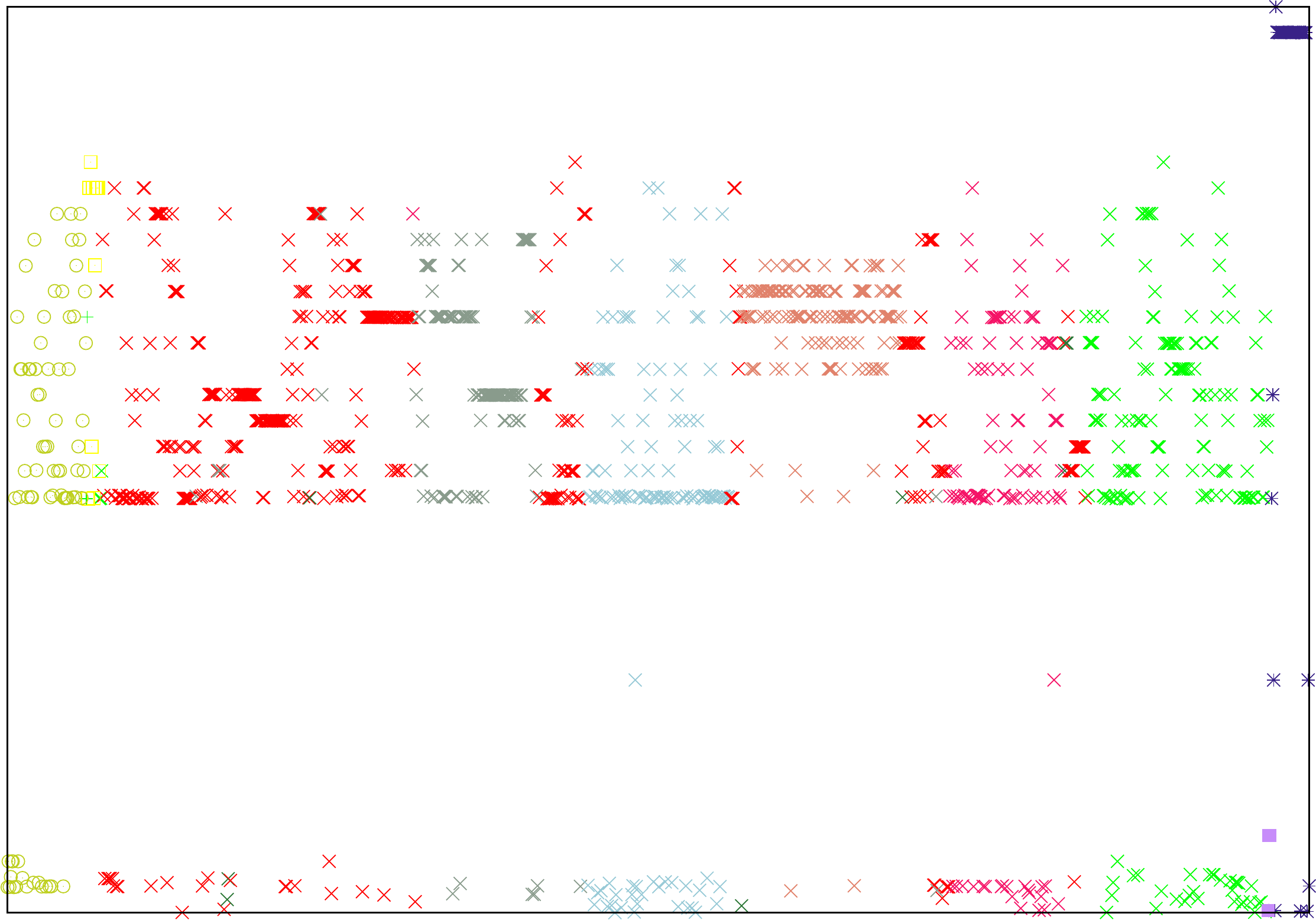} 
  \caption{An instance of \fsgenoma.  \label{fig:genoma}}
\end{figure}

Figure~\ref{fig:genoma} shows a graphical representation of a FSG of a specific Debian GNU/ Linux OS installation. 
In the figure, the x -axis lists the $120,000$ files that form the operating system installation, while the y-axis 
indicates the location of the first file block allocated for each file (the other file blocks are ignored in 
drawing the figure for clarity). We colored each point with the color of the parent directory to which the file 
belongs (in details, we consider only the /boot, /etc, /lib, /root, /usr and /var parent directories). As can be 
seen from the colors in the figure, the files in different parent directories tend to intermingle in the same 
areas of the storage device.

\subsubsection{Entropy and min-entropy\label{sec:exp-entropy}}

The entropy and the min-entropy are largely used in many fields of security and 
networking, as metrics to distinguish network attacks or connection 
type~\cite{entropynetwork,realdos}. In fact, entropy-based approaches are attractive
since they provide more fine-grained insights than statistical traffic analysis
techniques.
\begin{comment}
In other computer science research fields, the min-entropy is also used in the
context of randomness extractors. The extractors are functions aiming at
generating independent and randomly distributed output starting from a biased
and correlated source of information. In such cases, the min-entropy is used as
measure of the amount of randomness achieved in the worst case.
\end{comment}
In our context, the entropy-based approaches capture fine-grained patterns in 
the selection of the block locations.

The concept of \textit{Entropy} was introduced by Shannon in~\cite{Shannon1948}.
The classic definition says that entropy is a measure of the \textit{uncertainty
associated with a random variable}. 

The entropy $H(X)$ of a discrete random variable $X$ is defined by the formula:
$$  
H(X) = - \sum_i p_i \log_2 p_i
$$
where $p_i=P[X=i]$ is the probability that $X$ assumes the value $i$.

Min-entropy captures the amount of randomness that can be provided by a given
distribution. According with the definition in~\cite{entropy}, we compute the
min-entropy with the following formula:
$$
min\ entropy(X) = min_{ x \in X}\ log_2 \ \left( \frac{1}{Pr[X=x]}
\right)
$$
In other words, the min-entropy of a distribution is equal to the probability of
the most likely element in $X$ being drawn from $X$. As such, the min-entropy
can be considered as a lower bound for the Shannon entropy.

The min-entropy of \fsgenoma\ is computed by Algorithm~\ref{alg:min-entr}. The
array $\vec{I}$ contains all the installations in our dataset; the array
$\vec{F}$ contains the list of files of the current installation.
$\vec{L}_i^f$ is the list of the block locations, $(l_1^i,\ldots,l_k^i)$, used by file
$f$ in installation $i$. The matrix $\mathcal{H}$ contains partial results; the
$k$-th row stores the statistics of the file $k$ for all the installations:
each column $j$ stores the number of times that the block location $j$ has been assigned
to the file $f$. The first part of the algorithm is devoted to the computation
of
the matrix $\mathcal{H}$, then, we compute the min-entropy for each file of the
installation. It is important to notice that the probability of choosing a location
$j$ of a file $f$ is calculated by dividing $\mathcal{H}[f][j]$ by the number of
performed installation, instead of the number of block locations available in the disk.
Indeed, the number of block locations in an $80GB$ HDD is greater then the number of
performed installations. Therefore, we consider this estimation as a lower bound of
the real value of the min-entropy. Note that this analysis has been inspired 
by~\cite{accellerometro}, where authors evaluate whether an accelerometer sensor
can be considered as source of randomness, and by~\cite{browseruniq}, where the author
evidences how modern web browsers can be fingerprinted.

\SetCommentSty{textit} 
\LinesNumbered 
\begin{algorithm}[!h]
  \caption{Evaluation of the min-entropy of each file created during the
installation procedure}
  \label{alg:min-entr}
  \small
  %\dontprintsemicolon
  
  \KwIn{\\
    $\vec{F}$: The array of file IDs\\
    $\vec{I}$: The array of samples to analyze \\
    %$\mathcal{H}$: The $n \times m$ min-entropy matrix. $n$ equals to the
    $\mathcal{H}$: The \fsg\ matrix; initialized to all $0$\\
    %$x$: The number of experiments
  }
  \KwOut{The min-entropy associated to each file}
  
  \BlankLine
  \textbf{ComputeMinEntropy}($\vec{F}$,$\vec{I}$, $\mathcal{H}$)
  \BlankLine
  \Begin{
    \ForEach{ $i \in \vec{I}$ } {
      \ForEach{$f \in \vec{F}$} {
        $\vec{L}_i^f = \{l_1^i,\ldots,l_k^i \in f\}$\\ 
        \ForEach{$l \in \vec{L}_i^f$} {
            $\mathcal{H}[f][l] \gets \mathcal{H}[f][l] + 1$
        }           
      }     
    }
    \BlankLine
    \ForEach{$f \in \mathcal{H}$}{
        v $\gets \max(\mathcal{H}[f][j]), j=1,\ldots,m $\\
        $E[f] \gets -log\left(\dfrac{v}{|\vec{I}|}\right)$
    }   
    \Return{$\vec{E}$}  
  }
  \BlankLine
\end{algorithm}

The entropy is estimated in a similar way; the only difference with respect to
Algorithm~\ref{alg:min-entr} is in the lines 11 up to 14 that are replaced by the 
entropy estimation according to the previous formula.

\begin{figure}[!h]
  \centering
  \subfigure[Standard installation]{
    \label{img:installationminH}
    \includegraphics[angle=-90,width=.45\textwidth]{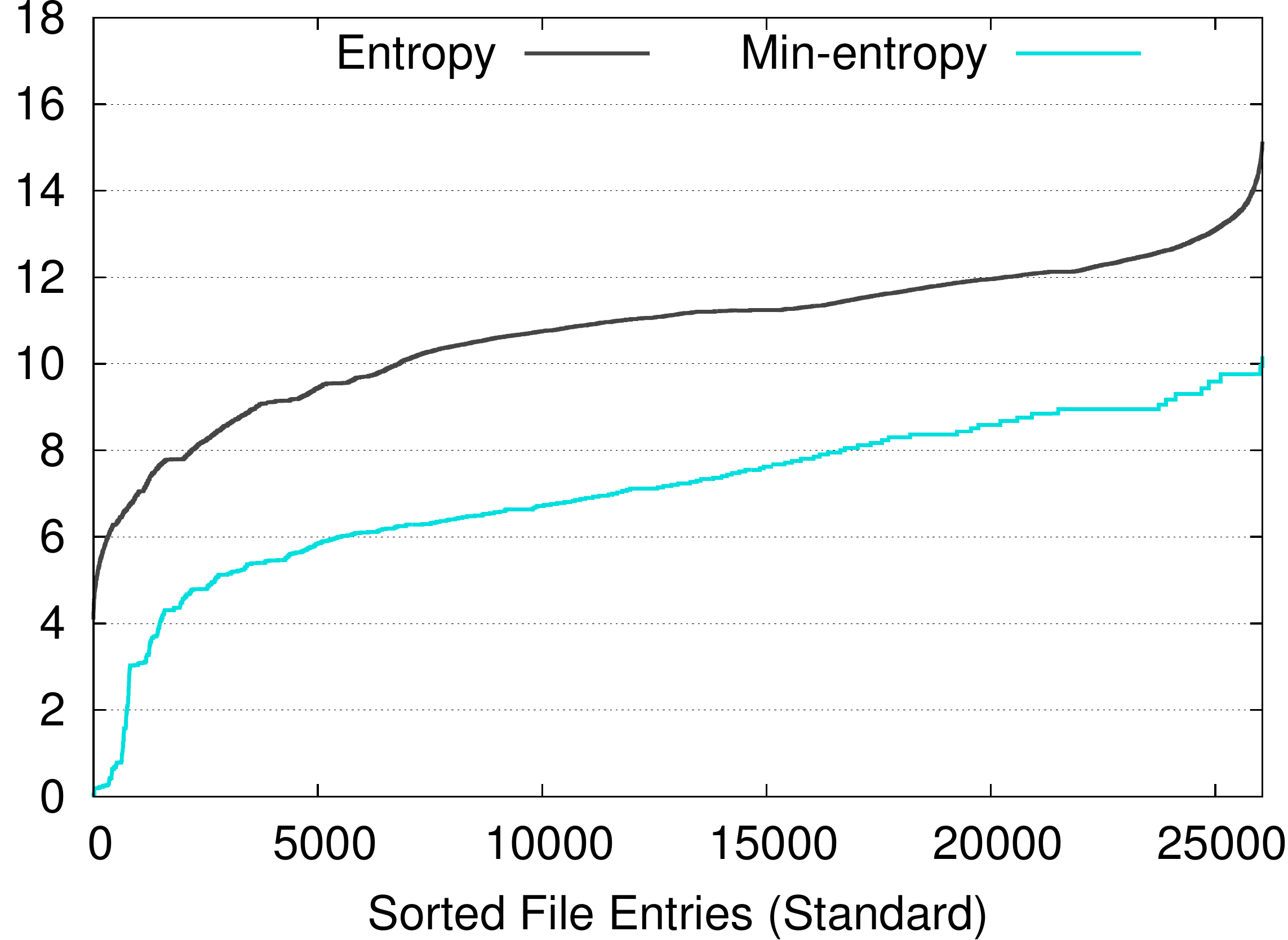}
  }
  \subfigure[Full Installation]{
    \includegraphics[angle=-90,width=.45\textwidth]{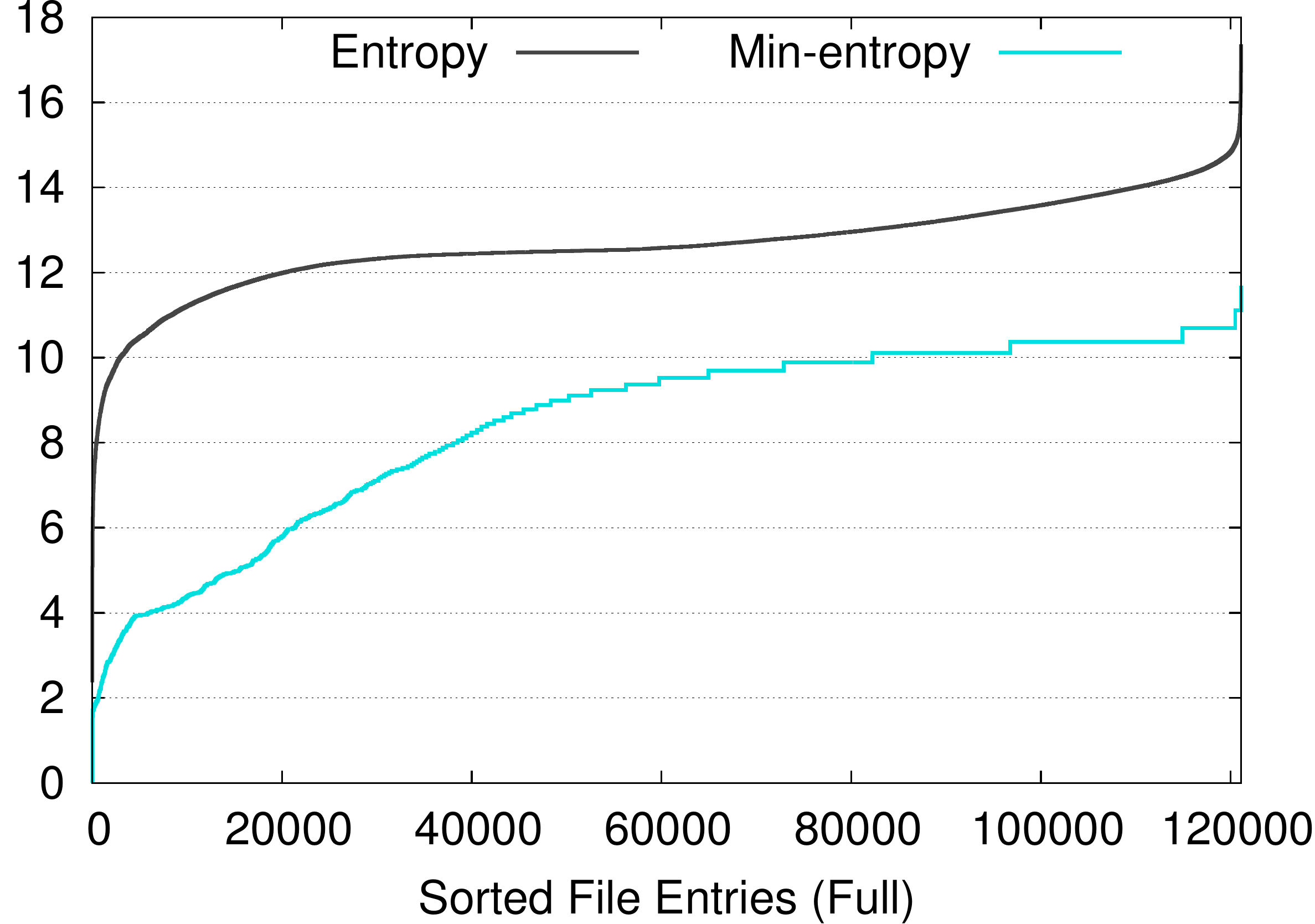}
    \label{img:syntheticminH}
  }
  \caption{Entropy and min-entropy estimation for the \genoma\ of 6.200
installation.}
\end{figure}

Figures~\ref{img:installationminH} and~\ref{img:syntheticminH} show the entropy
and min-entropy of the \textit{standard} and \textit{full} installations
respectively. The graphs represents the evaluation related to $6,200$ for each
of the installations types.
On the \textit{x}-axis are reported the files considered to 
evaluate entropy and min-entropy. Note that, the \textit{full} installation contains more 
files rather than the \textit{standard} one, i.e.
the \textit{x}-axes have well distinct order of magnitude. As the graphs show,
the operating system strives to cluster the blocks of the file---that is, exploiting the space locality
principle. 
Nonetheless, as it can be noticed, the entropy and min-entropy values reach high
values, justifying the rational of our proposal.
As such the results of the entropy measures highlight a significant amount of diversity in the collected 
corpus of installations and suggest that \fsgenoma\ can be considered a distinctive
and unique feature. 
Moreover, all the experiments have been performed using the same components (in terms of hardware
and software); we emphasize this is the worst case, in fact, using distinct
configuration (e.g. different HDD and CPU as well as \fs\ ) we can get even more 
profitable results.

\section{Related work\label{sec:related}}

The research area closest to the issues addressed in our paper appears to be
Forensic identification. It is worth describing some of these related results,
even though we stress that the approach we consider in this paper has
not been considered before.

Forensic identification is subject of many research and it is used in many
applications. A  well know forensic identification task involves the
characterization of printers. Despite the large use of digital communications,
printed document is still used in many criminal and malicious activities. 
Consider for examples the forgery or alteration of national documents, e.g. an
ID card. Printed material may be also used as evidence of illicit activities,
e.g. meeting notes, paper correspondences, boarding passes or bank
transactions,~\cite{printforensic}. Hence, it may be desirable to have the
capability of identifying the manufacturer, model, or even a specific device
that was used to produce a paper document. In this context, forensic
characterization of printer papers involves finding features in the printed
document that are uniquely related to a specific printer. 
This printer signatures directly emerge from printer mechanism as shown by many
research,~\cite{printsignature,printgauss} and others. Forensic identification
of printers are manifold. 
A forensic technique used for the identification of printer device exploits the
print quality defect, namely the banding in electro-photographic (EP). Results
in literature, show that different printers expose different sets of banding
frequencies that are related to brand and model.
In the same fashion, it is possible to extract printer signatures by capturing
the banding signal within the height of one text character. A surprising
result,~\cite{printsvm}, shows that the print quality defects, are considered as
texture in the printed regions of the document. The features can be extracted
from individual printed characters, in particular, the letter ‘‘e’’ in a
document. In this case, authors extract the signature using machine learning
approaches. 

Next, there exists many forensic identification technique for wired
and wireless (RF) communication devices. 
In~\cite{wiredforensic}, authors propose a novel approach based on the
utilization of the signal characteristics of wired Ethernet card for use in a
security context. In fact, the ability of distinguish two NIC may be used as
communication filter able to prevent arp spoofing attacks and, in general, to
control the authenticity of either the source or the destination of a wired
communications. The signature of a wired network interface can be extracted from
the minute variations of the transient portion extracted by the produced signal.
This variations result from hardware and manufacturing inconsistencies,

Wireless devices can be identified using similar approaches. Software that
implements such forensic mechanisms can remotely identify the types of devices
located in a specific environment, realizing a layer $1$ Intrusion Detection
Systems. These techniques,~\cite{rfmodulation}, are usually implemented by the
following steps: i) a specially designed probe signal is sent in the wireless
domain, ii) the response signal is then captured and iii) exterminated in order
to verify the identity of the wireless device. In fact, the response signal
contains unique distortion feature generated by the RF circuitry of the device
in a nonlinearities fashion; these distortions are used as fingerprint of the
device.
\begin{comment}
MIMMO: Questa parte accenna un related work per EXT4 senza entrare nei dettagli,
senza spiegare i vantaggi e i contributi di questo articolo. Lo ho tolto perche'
la parte sui FS la abbiamo messa prima!

Many researchs investigated the features of Ext4 for forensic
purposes~\cite{forensicext4,ext4performance}. However, none of the previous
works address the issue of the identification of the device through the analysis
of the FST. 
\end{comment}

\section{Conclusions\label{sec:conclusions}}

This paper introduces \fsgenoma, a novel solution for the unique identification of HDDs. 
The \fsgenoma\ technique leverages the diversity of the physical
blocks locations used to store the files of an operating system. 

To support our proposal, we justify our rationale, providing an insight of the
reasons that make two different installations of the same OS on the same machine
fundamentally different, when analyzed with the lens of our proposal. 
Our findings are supported by %we have performed 
an extensive experimental campaign: we
carried out some $12,000$ installations of the Debian GNU/Linux operating systems. All the installations have been performed in a real testbed using the
same hardware (storage devices, motherboard, CPU, etc..). The resulting huge data set has been analyzed resorting to sound information theory tools (entropy and min-entropy).
All the analyzed metrics show 
that the File System Genome is a distinctive
and unique feature of each individual HDD. 
Note that, 
given the rationales our solution is based upon, 
we expect that the \fsgenoma\ can be successfully applied to other mass storage devices as well (e.g. SSDs, Flash memories).

As for future work, we plan to spend %In the short term, 
our efforts spent in %are directed towards 
two directions: 
the use of
the \fsgenoma\ as a building block for authentication protocols based on challenge/response; and,
the definition of the  genomes of other species,
namely other file systems and operating systems.

\bibliographystyle{plain}
\bibliography{osfingerprint}

\begin{thebibliography}{10}

\bibitem{printsignature}
G.~N. Ali, P.~Chiang, A.K. Mikkilineni, J.P. Allebach, G.T.C. Chiu, and E.J.
  Delp.
\newblock Intrinsic and extrinsic signatures for information hiding and secure
  printing with electrophotographic devices.
\newblock In {\em International Conference on Digital Printing Technologies},
  pages 511--515, 2003.

\bibitem{printgauss}
Gazi~N. Ali, Pei-Ju Chiang, Aravind~K. Mikkilineni, George~T.C. Chiu, Edward~J.
  Delp, and Jan~P. Allebach.
\newblock Application of principal components analysis and gaussian mixture
  models to printer identification.
\newblock In {\em International Conference on Digital Printing Technologies},
  volume~20, pages 301--305, 10 2004.

\bibitem{securedigitalcamera}
Paul Blythe and Jessica Fridrich.
\newblock Secure digital camera.
\newblock In {\em in Proceedings of Digital Forensic Research Workshop}, pages
  17--19, 2004.

\bibitem{understanding}
Daniel Bovet and Marco Cesati.
\newblock {\em Understanding the Linux Kernel, Second Edition}.
\newblock O'Reilly \& Associates, Inc., Sebastopol, CA, USA, 2 edition, 2002.

\bibitem{browseruniq}
Peter Eckersley.
\newblock How unique is your web browser?
\newblock In {\em Proceedings of the 10th international conference on Privacy
  enhancing technologies}, pages 1--18, 2010.

\bibitem{wiredforensic}
R.M. Gerdes, M.~Mina, S.F. Russell, and T.E. Daniels.
\newblock Physical-layer identification of wired ethernet devices.
\newblock In {\em Information Forensics and Security, IEEE Transactions on},
  volume~7, pages 1339--1353, 2012.

\bibitem{entropynetwork}
Yu~Gu, Andrew McCallum, and Don Towsley.
\newblock Detecting anomalies in network traffic using maximum entropy
  estimation.
\newblock In {\em In ACM SIGCOMM}, pages 32--32, 2005.

\bibitem{survey-forensics}
Nitin Khanna, Aravind~K. Mikkilineni, Anthony~F. Martone, Gazi~N. Ali, George
  T.~C. Chiu, Jan~P. Allebach, and Edward~J. Delp.
\newblock A survey of forensic characterization methods for physical devices.
\newblock In {\em Digital Investigation.}, volume~3, pages 17--28, 2006.

\bibitem{ext4}
A.~Mathur, M~Cao, S.~Bhattacharya, A.~Dilger, A.~Tomas, and L.~Vivier.
\newblock {The new ext4 filesystem: current status and future plans}.
\newblock In {\em Linux Symposium}, 2007.

\bibitem{printforensic}
Aravind~K. Mikkilineni, Gazi~N. Ali, Pei-Ju Chiang, George T.-C. Chiu, Jan~P.
  Allebach, and Edward~J. Delp.
\newblock Signature-embedding in printed documents for security and forensic
  applications.
\newblock In {\em Security, Steganography, and Watermarking of Multimedia
  Contents}, pages 455--466, 2004.

\bibitem{printsvm}
Aravind~K. Mikkilineni, Osman Arslan, Pei ju~Chiang, Roy~M. Kumontoy, Jan~P.
  Allebach, and George~T. c.
\newblock Printer forensics using svm techniques.
\newblock In {\em International Conference on Digital Printing Technologies},
  pages 223--226, 2005.

\bibitem{rfmodulation}
J.C. Pedro and N.B. Carvalho.
\newblock {\em Intermodulation Distortion in Microwave and Wireless Circuits}.
\newblock Artech House Microwave Library. Artech House, 2003.

\bibitem{entropy}
A.~Renyi.
\newblock {On measures of information and entropy}.
\newblock In {\em Proceedings of the 4th Berkeley Symposium on Mathematics,
  Statistics and Probability}, pages 547--561, 1960.

\bibitem{Shannon1948}
Claude~E. Shannon.
\newblock A mathematical theory of communication.
\newblock In {\em The Bell system technical journal}, volume~27, pages
  379--423, 1948.

\bibitem{realdos}
Domenico Vitali, Antonio Villani, Angelo Spognardi, Roberto Battistoni, and
  Luigi~V. Mancini.
\newblock Ddos detection with information theory metrics and netflows - a real
  case.
\newblock In {\em SECRYPT}, pages 172--181, 2012.

\bibitem{accellerometro}
J.~Voris, N.~Saxena, and T.~Halevi.
\newblock Accelerometers and randomness: perfect together.
\newblock In {\em Proceedings of the fourth ACM conference on Wireless network
  security}, pages 115--126, 2011.

\end{thebibliography}

\end{document}